# Imaging Andreev Reflection in Graphene

S. Bhandari,[†,‡] G.-H. Lee,[†,§] K. Watanabe,[¶] T. Taniguchi,[¶] P. Kim[†,&] and R. M. Westervelt[†,&,*]

[†] School of Engineering and Applied Sciences, Harvard University, Cambridge MA, 02138 USA

[‡] Dept. of Physics and Engineering, Slippery Rock University, Slippery Rock PA 16057, USA

[§] Dept. of Physics, Pohang University of Science and Technology, Pohang, Republic of Korea

[¶] National Institute for Materials Science, 1-1 Namiki, Tsukuba 305-0044, Japan

[&] Dept. of Physics, Harvard University, Cambridge MA, 02138, USA

*Email: westervelt@seas.harvard.edu

**ABSTRACT:** Coherent charge transport along ballistic paths can be introduced into graphene by Andreev reflection, for which an electron reflects from a superconducting contact as a hole, while a Cooper pair is transmitted. We use a liquid-helium cooled scanning gate microscope (SGM) to image Andreev reflection in graphene in the magnetic focusing regime, where carriers move along cyclotron orbits between contacts. Images of flow are obtained by deflecting carrier paths and displaying the resulting change in conductance. When electrons enter the the superconductor, Andreev-reflected holes leave for the collecting contact. To test the results, we destroy Andreev reflection with a large current and by heating above the critical temperature. In both cases, the reflected carriers change from holes to electrons.

Electrons in graphene have remarkable characteristics that pave the way for ballistic electronic devices.[1,2] Coherent charge transport can be introduced into graphene from superconductors by Andreev scattering[3-6] or Josephson coupling.[7-14] Andreev reflection allows an electron to reflect from a superconductor as a hole, while a Cooper pair is transmitted. Graphene/superconductor hybrid devices show coherent phenomena including crossed Andreev conversion[7] and edge states in graphene Josephson junctions.[8] Here, we employ a liquid-helium cooled scanning gate microscope (SGM)[15-26] to image Andreev reflection in graphene from a superconducting contact in the magnetic focusing regime. The SGM images the ballistic paths of electrons or holes by deflecting their trajectories and displaying the resulting change in conductance.[19,21] The images show cyclotron orbits of electrons entering the superconductor and Andreev-reflected holes leaving for the collecting contact. To confirm the results, we destroy Andreev reflection by applying a large current, and by heating above the critical temperature. For both cases, the collected carriers change from holes to electrons.

Andreev reflection[3-5] is the process that links carriers in graphene with a superconducting contact – Fig. 1(a) shows how this occurs. An electron enters the superconducting contact at an energy $E_F$ + e$V_{bs}$, where e$V_{bs}$ is less than the superconducting energy gap $\Delta$. A hole is reflected back into the graphene with energy $E_F$ - e$V_{bs}$, as well as a Cooper pair that passes into the superconductor. Energy, momentum and charge are conserved. These processes are indicated on the graphene band structure $E$ *vs.* $k$ shown on the right – note that the electron and Andreev-reflected hole are both near $E_F$ in the conduction band.

Andreev reflection is a microscopic description of the superconducting proximity effect and explains how non-superconducting charge carriers gain the superconducting correlation. This process is essential for superconducting hybrid systems, including superconducting quantum circuits, Josephson junctions, and topological superconductivity. A direct spatial mapping of the Andreev process in a ballistic conductor can help to understand the microscopic details of superconducting proximity effect in the hybrid systems.

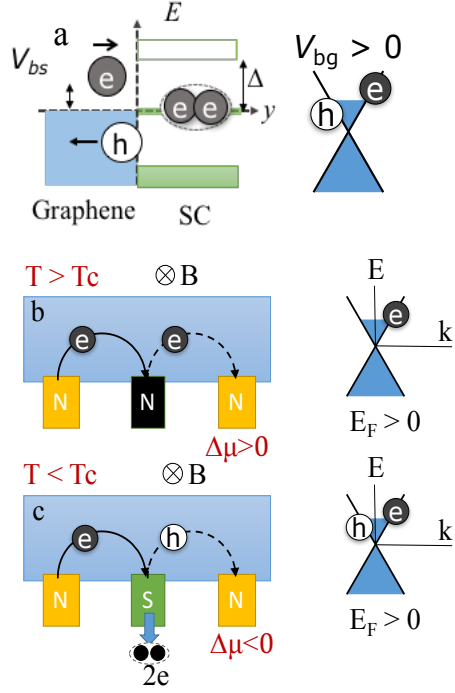

**Figure 1:** Andreev-reflection in a graphene device with superconducting contacts. In the following, $E_F$ is the Fermi energy, $\Delta$ is the superconducting energy gap, and $e$ is the electronic charge. (a) For Andreev reflection, an electron with energy $E_F + eV_{bs}$ entering a superconducting contact with $eV_{bs} < \Delta$, generates a hole with energy $E_F - eV_{bs}$ that is reflected back into the graphene, and a Cooper pair that passes into the superconductor. The dispersion relation $E$ vs. $k$. of graphene is shown on the right. (b) Illustration of magnetic focusing of electron orbits for a graphene sample with normal contacts in a perpendicular magnetic field $B$ when the cyclotron diameter matches the contact spacing. (c) For a superconducting center contact, Andreev reflection converts an incoming electron into an outgoing hole with positive charge that passes onto the right contact, and a Cooper pair flowing into the superconductor.

Here, we use magnetic focusing to track the motion of electrons and Andreev-reflected holes through a graphene device with transparent niobium (Nb) superconducting contacts, as illustrated in Figs. 1(b) and 1(c). A perpendicular magnetic field $B$ bends carrier motion into circular cyclotron orbits with diameter $d_c = (n/\pi)^{1/2} h/eB$, where $n$ is the carrier density, and $e$ is the elementary charge. A magnetic focusing peak occurs when carriers leaving the first contact are rejoined at a second contact spaced a distance $d_c$ away. Figure 1(b) illustrates the usual reflection of electrons that occurs from a normal center contact in the magnetic focusing regime: an electron impinging on the contact is reflected as an electron and travels along a cyclotron orbit to the right contact. Figure 1(c) shows how Andreev reflection occurs for a superconducting center contact. In this case, an electron entering the center contact is reflected as a hole with the opposite charge. The hole travels along the same cyclotron orbit as the electron to the right contact, because both particles are in the conduction band. An immediate indicator for Andreev reflection is that the voltage signal on the right contact has reversed sign, as shown below.

Figure 2 shows a scanning electron micrograph of the graphene device, fabricated from a hexagonal boron nitride (hBN) encapsulated monolayer graphene sheet with a 35 nm thick hBN layer to top and a 45 nm hBN layer on bottom, placed on an Si substrate with 300 nm thick $SiO_2$ layer, which acts as a back gate. The encapsulated sheet was shaped into a Hall bar with five Nb superconducting contacts S1, S2, S3, S4 and S5, with critical temperature $T_c = 8.0$ K. The sample was mounted inside the cooled scanning gate microscope (SGM) in an inner chamber filled with He-4 gas at 3 Torr surrounded by liquid He at 4.2K. A perpendicular magnetic field $B$ was applied using a superconducting solenoid. The semi-circles in Fig. 2a illustrate the expected cyclotron orbits of electrons (blue) and Andreev-reflected holes (red). To record an image of electron flow, a current $I$ is passed into the device from S2 while S1 and S3 are grounded, and the voltage difference $V_m$ between S4 and S5 is measured as the tip is raster scanned across the sample. Such non-local measurements can avoid unwanted background signals. The transmission $T_m$ of electrons (or holes) from S2 to S4 is proportional to the signal $V_m = IR_m$, where $R_m$ is transresistance - this happens because the voltmeter draws no current. As electrons flow into S4, the electron density in the contact increases. The resulting increase in chemical potential $\Delta\mu$ drives a reverse flow of electrons that is sufficient to zero the total current into S4. The resulting voltage $V_m \sim -\Delta\mu$ is negative, because electrons are negatively charged. If holes enter S4, the electron density and chemical potential decrease instead, and the signal $V_m$ is positive. This change of sign allows us to determine whether electrons or holes are entering the collecting contact S4, as demonstrated below.

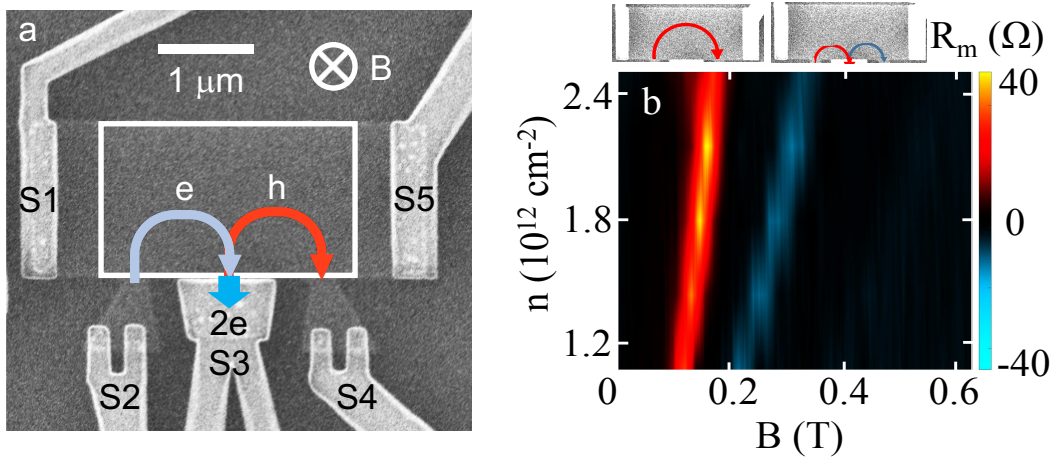

**Figure 2:** Magnetic focusing of Andreev reflection in graphene. (a) Scanning electron micrograph of the hBN encapsulated graphene device on an Si/SiO$_2$ substrate with superconducting contacts S1, S2, S3, S4 and S5. The blue and red semi-circles illustrate the cyclotron orbits of an electron and an Andreev-reflected hole in a perpendicular magnetic field $B$. A current $I$ enters the device from S2 while S1 and S3 are grounded and the voltage $V_m$ between S4 and S5 is measured. (b) Measured transresistance $R_m = V_m/I$ is displayed *vs.* $B$ and electron density $n$ at 4.2 K. The first magnetic focusing peak (red) between the outer contacts S2 and S4 occurs when the electron cyclotron diameter matches their separation. A second magnetic focusing peak (blue) of opposite sign occurs, when electrons from S2 travel along a cyclotron orbit to S3 and are Andreev-reflected as holes that follow an orbit to S4.

To determine the magnetic fields $B$ and electron gas density $n$ at which magnetic focusing occurs between the contacts, the measured transresistance $R_m$ is displayed *vs.* $B$ and backgate-tuned $n$ at 4.2 K in Fig. 2(b). The first 'peak' (red) corresponds to magnetic focusing of electrons between the two outer contacts S2 and S4, as illustrated in the inset to Fig. 2(b) . At larger $B$, a second 'peak' (blue) corresponds to magnetic focusing of Andreev-reflected holes between the center and right contacts, S3 and S4, also illustrated in the inset to Fig. 2(b). The sign of the signal $R_m$ reverses, indicating that the carriers arriving at S4 now have positive charge.

We use a cooled SGM to image the flow of electrons and Andreev-reflected holes through the graphene device on the two magnetic focusing peaks in $B$ and $n$ shown in Fig. 2(b). The imaging technique was described in detail for our previous imaging experiments on graphene.[25,26] A charged tip is scanned at a constant height above the graphene device, creating an image charge below the tip in the two-dimensional electron gas (2DEG) that deflects electrons away from their original trajectories. In consequence, the transmission of electrons (or holes) from an emitting to a collecting contact is reduced. An image of electron (or hole) flow is obtained by displaying the measured change $\Delta R_m$ as the tip is raster scanned across the sample, where $\Delta R_m$ is proportional to the change in carrier transmission $\Delta T_m$.[25,26]

In our earlier imaging work on graphene,[25,26] we gave a detailed description how to use ray tracing to simulate SGM images of carrier flow. Electrostatics provides a formula for the spatial profile of the image charge in the 2DEG beneath the charged SGM tip, which produces a local dip in electron density of radius 70 nm comparable to the height of the tip above the graphene layer.

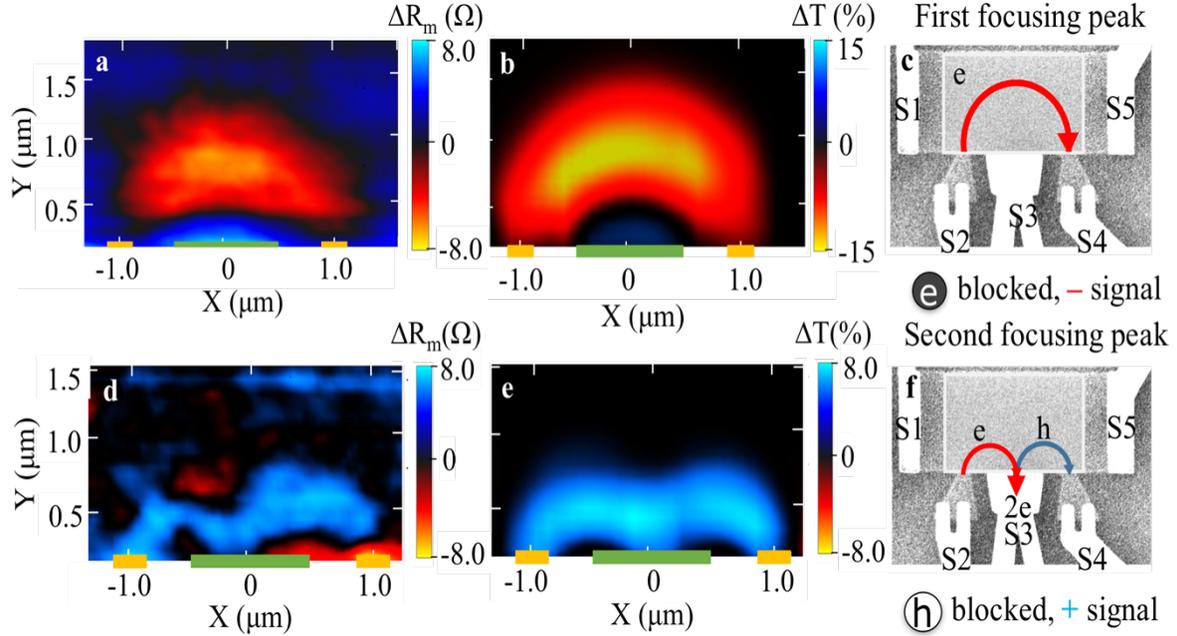

**Figure 3:** Images of Andreev reflection in the graphene from a superconducting contact. (a) SGM image at the first magnetic focusing peak showing cyclotron orbits of electrons (red region) from contact S2 to S4, taken at $B = 0.13$ T, $n = 1.8 \times 10^{12}$ cm$^{-2}$ and 4.2 K. (b) Corresponding simulated image. (c) Illustration of electron cyclotron orbits (red) between contacts S2 to S4. The SGM tip blocks electron orbits and decreases their transmission $T$. (d) SGM image of Andreev reflection taken at the second magnetic focusing peak, which shows two cyclotron orbits (blue region), one for electrons from contact S2 to S3, and one for Andreev-reflected holes from contact S3 to S4, taken at $B = 0.26$ T. The signal reverses sign, because holes are arriving at S4, not electrons. (e) Corresponding simulated image. (f) Illustration of the electron orbit from contact S2 to S3, and the Andreev reflected hole orbit from S3 to S4. Both orbits appear blue in the SGM image, because the electron orbit between contacts S2 and S3 was blocked, stopping the Andreev reflected hole that would have traveled to contact S4.

The density dip beneath the tip depresses the local chemical potential, which is the Fermi energy $E_F$. In turn, the diffusive motion of electrons set up by the change in $E_F$ tries to fill the dip. In balance, the total chemical potential $E_F(\vec{r}) + U(\vec{r})$ is constant, where $U(\vec{r})$ is the potential energy of an electron, and the force on a nearby electron is $\vec{F}(\vec{r}) = -\vec{\nabla}U(\vec{r})$. Ray tracing simulations of the transmission $T$ between two contacts are carried out by starting a large number of trajectories at random angles from the emitting contact, calculating their trajectories through the device using the classical equation of motion, and counting the fraction of these orbits that reach the collecting contact.[25,26]

The SGM images in Fig. 3 clearly demonstrate how the transition to Andreev reflection occurs. The top three panels show the flow of electrons along a cyclotron orbit between the two outer contacts S2 and S4 on the first magnetic focusing peak for $B$ = 0.13 T at 4.2 K. Figure 3(a) presents an SGM image of electron flow, Fig. 3(b) shows corresponding simulations, which are in good agreement, and Fig. 3(c) illustrates the electron orbit from S2 to S4 in a scanning electron micrograph of the device. The orbits are shown red ($\Delta R_m < 0$) in Fig. 3(a), because negative charges – electrons – travel along the orbit and enter contact S4.

The lower three panels in Fig. 3 show Andreev reflection patterns of carrier flow, taken on the second magnetic peak in Fig. 2(b). Figure 3(d) presents an SGM image of the flow of electrons from contact S2 that follow a cyclotron orbit to the superconducting center contact S3, as well as the Andreev-reflected holes that leave S3 and follow a cyclotron orbit to the collecting contact S4. Both orbits are shown blue ($\Delta R_m > 0$), indicating that positively charged carriers - holes - enter S4. The initial electron orbit from S2 to S3 is also blue, because those electrons have been Andreev-reflected as holes. Charge is conserved, because two electrons simultaneously pass into the superconducting contact S3 as a Cooper pair, then flow to ground. Figure 3(f) shows ray-tracing simulations of Andreev reflection that are in good agreement, and Fig. 3(g) illustrates both the initial electron cyclotron orbit from S2 to S3, and the Andreev-reflected hole orbits from S3 to S4 on the device. The use of magnetic focusing allows us to clearly identify the carrier flow associated with Andreev reflection. We expect our SGM imaging technique to also be useful to characterize Andreev reflection in other circumstances where a magnetic field is not present.

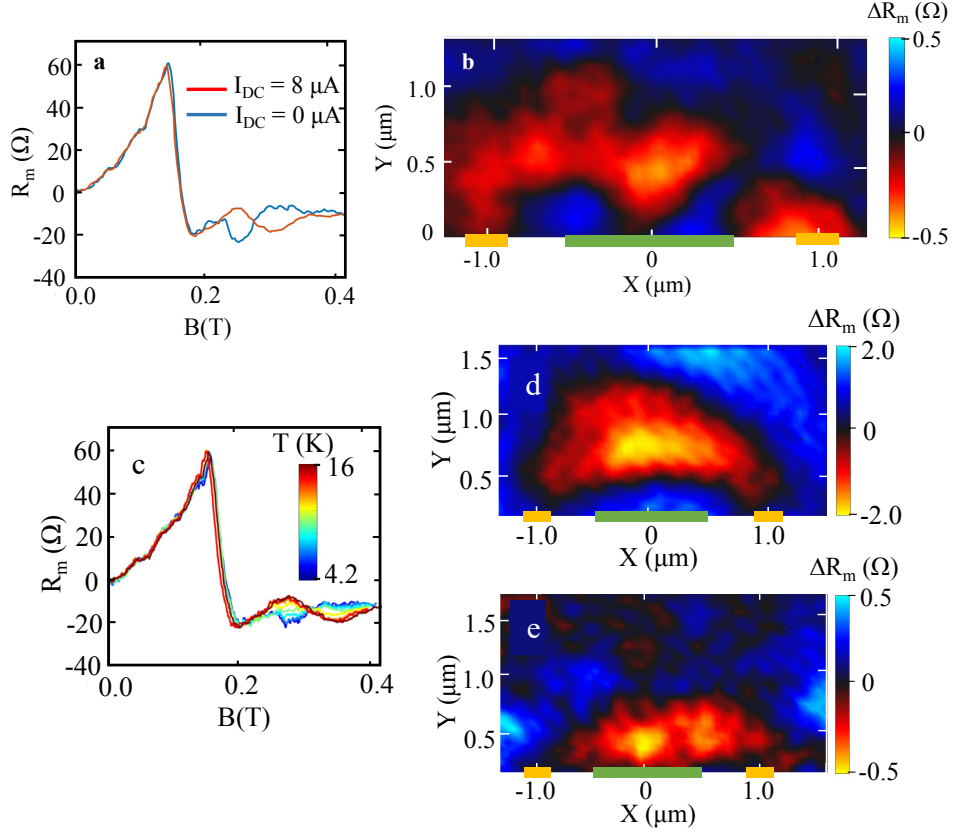

**Figure 4:** Destruction of Andreev reflection by a large current and by heating above the critical temperature $T_c$. (a) Plot of $R_m$ vs. $B$ at $n = 1.8 \times 10^{12}$ cm$^{-2}$ shows that the magnetic focusing dip (blue) at $B = 0.26$ T, for Andreev reflected holes arriving at contact S4 for $I_{DC} = 0$ has been transformed into a peak (red) showing that electrons arrive instead for $I_{DC} = 8$ µA. (b) An SGM image of carrier flow at high current is red ($\Delta R_m < 0$) showing that electrons flow into the right contact S4 instead of holes. (c) Increasing the temperature from 4.2 K to 16 K at $n = 1.8 \times 10^{12}$ cm$^{-2}$ changes the Andreev reflection magnetic focusing dip in $R_m$ at $B = 0.26$ T, for which holes enter contact S4, to a peak that is appropriate for electrons. (d) SGM image of carrier flow taken at 16 K $> T_c$ shows electrons following cyclotron orbits (red) from S2 to S4 on the first magnetic focusing peak for $B = 0.13$ T. (e) SGM image at the second magnetic focusing peak for temperatures above $T_c$ shows that the hole signal (blue) observed below $T_c$ in Fig. 3d has changed into electrons (red) as superconductivity goes away.

To verify that Andreev reflection causes the patterns of carrier flow observed Fig. 3 we apply an emitter current $I_{DC} = 8$ µA for which the emitted electron energy $eV_{bs} > \Delta$ and does not undergo Andreev reflection at the superconducting contact S3. In Fig. 4(a), the dip in $R_m$ for $I_{DC} = 0$ (blue) at the second magnetic focusing peak that is indicative of Andreev reflection turns into a local maximum for $I_{DC} = 8$ µA (red), showing that the carriers received by contact S4 have changed from holes to electrons. In addition, the SGM image of carrier flow in Fig. 4(b) for $I_{DC} = 8$ µA shows that the charge carriers have changed sign, where red ($\Delta R_m < 0$) replaces the blue ($\Delta R_m > 0$) regions in Fig. 3(d), because electrons are collected by contact S4 instead of holes.

We carried out an additional test by heating the device above the critical temperature $T_c = 8.0$ K of the superconducting contact S3 to destroy Andreev reflection. Figure 4(c) plots the magnetic focusing signal $R_m$ vs. $B$ at a temperature 4.2 K, below $T_c$ (red), and at 16 K, above $T_c$ (blue). The dip in $R_m$ at $B = 2.6$ T is a signature of Andreev reflection, on the second magnetic focusing peak shown in Fig. 2(b). When the sample is warmed to 16 K, above $T_c$, the dip reverses sign to become a peak, suggesting that normal electron reflection occurs instead. In addition, Fig. 4(d) presents SGM images at 16 K. On the first magentic focusing peak at $B = 0.13$ T, electron cyclotron orbits that connect the two outer contacts S2 and S4 as shown in Fig. 3(a), because Andreev reflection is not involved. In stark contrast, SGM images of carrier flow in Fig. 4(e) for the second focusing peak at B = 2.6 T have changed sign from blue ($\Delta R_m > 0$) to red ($\Delta R_m < 0$) as the device is warmed from 4.2 K to 16 K. These results confirm that Andreev reflection has been destroyed and that electrons now enter the collecting contact S4 instead of holes.

**Acknowledgement.** The imaging research by S.B. and R.M.W. was supported by the DOE Basic Energy Sciences under grant DE-FG02-07ER46422. Sample fabrication and chaterization by G.H.L. and P.K was supported by the DOE Basic Energy Sciences under grant DE-SC0012260. The growth of hexagonal boron nitride crystals by K.W. and T.T. was supported by the Elemental Strategy Initiative conducted by MEXT, Japan, and CREST(JPMJCR15F3), JST.

**References**

(1) Geim, A., & Novoselov, K. The rise of graphene. *Nature Mater*. **6,** 183-191 (2007).

(2) Dean, C.R., Young, A.F., Meric, I., Lee, C., Wang, L., Sorgenfrei, S., Watanabe, K., Taniguchi, T., Kim, P., Shephard, K.L., & Hone, J. Boron nitride substrates for high-quality graphene electronics. *Nat. Nanotechnol.* **5,** 722-726 (2010).

(3) Andreev, A. F. The Thermal Conductivity of the Intermediate State in Superconductors. *Sov. Phys. JETP* **19**, 1228 (1964).

(4) Beenakker C.W.J. Specular Andreev reflection in graphene. *Phys. Rev. Lett.* **97**, 067007 (2006).

(5) Beenakker C.W.J. Colloquium: Andreev reflection and Klein tunnelling in graphene. *Rev. Mod. Phys*. **80**, 1337-1354 (2008).

(6) Efetov, D.K., Wang, L., Handschin, C., Efetov. K.B., Shuang, J., Cava, R., Taniguchi, T., Watanabe, K., Hone, J., Dean, C.R., & Kim, P. Specular interband Andreev reflections at van der Waals interfaces between graphene and $NbSe_2$. *Nature Phys*. **12**, 328-332 (2016).

(7) Lee G.-H., Huang, K.-F., Efetov, D.K., Wei, D.S., Hart, S., Taniguchi, T., Watanabe, K., Yacoby, A., & Kim, P. Inducing superconducting correlation in quantum Hall edge states. *Nature Phys*. **13**, 693-698 (2017).

(8) Allen, M.T., Shtanko, S., Fulga, I.C., Akhmerov, A.R., Watanabe, K., Taniguchi, T., Jarillo-Herrero, P., Levitov, L.S., & Yacoby, A. Spatially resolved edge currents and guided wave electronic states in graphene. *Nature Phys*. **12,** 128-133 (2016).

(9) Heersche, H.B., Jarillo-Herrero, P., Oostinga, J.B., Vandersypen, L.M.K., & Morpurgo, A.F. Bipolar supercurrent in graphene. *Nature* **446** 56-59 (2007).

(10) Mizuno, N., Nielson, B., & Du, X. Ballistic-like supercurrent in suspended graphene Josephson weak links. *Nat. Commun*. **4** 2716 (2013).

(11) Lee, G.-H., Kim, S., Jhi, S.-H., & Lee, H.J. Ultimately short ballistic vertical graphene Josephson junctions. *Nat. Commun*. **6**, 6181 (2015).

(12) Calado, V.E., Goswami, S., Nanda, G., Diez, M., Akhmerov, A.R., Watanabe, K., Taniguchi, T., Klapwijk, T.M., & Vandersypen, L.M.K. Ballistic Josephson junctions in edge-contacted graphene. *Nat. Nanotechnol.* **10**, 761-764 (2015).

(13) Ben Shalom, M., Zhu, M.J., Fal'ko, V.I., Mishchenko, A., Kretinin, A.V., Novoselov, K.S., Woods, R., Watanabe, K., Taniguchi, T., & Geim, A.K. Quantum oscillations of the critical current and high-field superconducting proximity in ballistic graphene. *Nature Phys*. **12**, 318-322 (2015).

(14) Amet, F., Ke, C.T., Borzenets, I.V., Wang, J., Watanabe, K., Taniguchi, T., Deacon, R.S., Yamamoto, M., Bomze, Y., Tarucha, S., & Finkelstein, G. Supercurrent in the quantum Hall regime. *Science* **352**, 966-969 (2016).

(15) Tessmer, S. H., Glicofridis, P. I., Ashoori, R. C., Levitov, L. S. & Melloch, M. R. Subsurface charge accumulation imaging of a quantum Hall liquid. *Nature* **392**, 51–54 (1998).

(16) Yacoby, A., Hess, H. F., Fulton, T. A., Pfeiffer, L. N. & West, K. W. Electrical imaging of the quantum Hall state. *Solid State Commun.* **111**, 1–13 (1999).

(17) McCormick, K. L. Woodside, M.T., Huang, M., Wu, M., McEuen, P.L., Duruoz, C., & Harris, J.S., Jr. Scanned potential microscopy of edge and bulk currents in the quantum Hall regime. *Phys. Rev. B* **59**, 4654–4657 (1999).

(18) Crook, R., Smith, C. G., Simmons, M. Y. & Ritchie, D. A. Imaging cyclotron orbits and scattering sites in a high-mobility two-dimensional electron gas. *Phys. Rev. B* **62**, 5174–5178 (2000).

(19) Topinka, M.A., Leroy, B.J., Shaw, S.E.J., Fleischmann, R., Heller, E.J., Westervelt, R.M., Maranowski, K.D., & Gossard, A.C. Imaging Coherent Electron Flow from a Quantum Point Contact. *Science* **289**, 2323-2326 (2000).

(20) Pioda, A., Kičin, S., Ihn, T., Sigrist, M., Fuhrer, A., Ensslin, K., Weichselbaum, A., Ulloa, S.E., Reinwald, M., & Wegscheider, W. Spatially Resolved Manipulation of Single Electrons in Quantum Dots Using a Scanned Probe. *Phys. Rev. Lett.* 93, 216801 (2004).

(21) LeRoy, B.J., Bleszynski, A.C., Aidala, K.E., Westervelt, R.M., Kalben, A., Heller, E.J., Shaw, S.E.J., Maranowski, K.D., & Gossard, A. C. Imaging Electron Interferometer. *Phys. Rev. Lett.* **94**, 126801 (2005).

(22) Steele, G. A., Ashoori, R. C., Pfeiffer, L. N. & West, K. W. Imaging transport resonances in the quantum Hall effect. *Phys. Rev. Lett.* **95**, 136804 (2005).

(23) Jura, M.P., Topinka, M.A., Urban, L., Yazdani, A., Shtrikman, H., Pfeiffer, L.N., West, K.W., & Goldhaber-Gordon, D. Unexpected features of branched flow through high-mobility two-dimensional electron gases. *Nature Phys*. **3**, 841 (2007).

(24) Aidala, K.E., Parrott, R.E., Kramer, T., Heller, E.J., Westervelt R.M., Hanson, M.P., & Gossard, A.C. Imaging magnetic focusing of coherent electron waves. *Nature Phys*. **3**, 464-468 (2007).

(25) Bhandari, S., Lee, G.-H., Klales, A., Watanabe, K., Taniguchi, T., Heller, E.J., Kim, P., & Westervelt, R.M. Imaging Cyclotron Orbits of Electrons in Graphene. *Nano Lett*. **16** 1690-1694 (2016).

(26) Bhandari, S., Lee, G.-H., Watanabe, K., Taniguchi,T., Kim, P., Westervelt, R.M. Imaging Electron Flow from Collimating Contacts in Graphene. *2D Materials* **5**(2), 021003 (2018).